\documentclass[aps,preprint]{revtex4}%
\usepackage{amsfonts}
\usepackage{amsmath}
\usepackage{amssymb}
\usepackage{graphicx}%
\providecommand{\U}[1]{\protect \rule{.1in}{.1in}}

\begin{document}
\preprint{ }
\title[FEM with spectral DVR basis functions]{Accuracy of a new hybrid finite element method for solving a scattering
Schr\"{o}dinger equation.}
\author{Joseph Power}
\affiliation{Physics Department, University of Connecticut, Storrs CT 06269}
\author{George Rawitscher}
\affiliation{Physics Department, University of Connecticut, Storrs CT 06269}
\pacs{PACS number}

\begin{abstract}
This hybrid method (FE-DVR), introduced by Resigno and McCurdy, Phys. Rev. A
\textbf{62}, 032706 (2000), uses Lagrange polynomials in each partition,
rather than \textquotedblleft hat\textquotedblright \ functions or Gaussian
functions. These polynomials are \ discrete variable representation functions,
and are.orthogonal under the Gauss-Lobatto quadrature discretization
approximation. Accuracy analyses of this method are performed for the case of
a one dimensional Schr\"{o}dinger equation with various types of local and
nonlocal potentials for scattering boundary conditions. The accuracy is
ascertained by comparison with a spectral Chebyshev integral equation method,
accurate to $1:10^{-11}$. For an accuracy of the phase shift of $1:10^{-8}$
The FE-DVR method is found to be $100$ times faster than a sixth order finite
difference method (Numerov), is easy to program, and can routinely achieve an
accuracy of better than $1:10^{-8}$ for the numerical examples studied.

\end{abstract}
\startpage{1}
\endpage{102}
\maketitle

\section{Introduction}

The solution of differential equations by means of expansions into discrete
variable representation (DVR) basis functions has become very popular since it
was first introduced in the early 1960's \cite{DVR1}. A review can be found in
the paper by Light and Carrington \cite{DVR2}, and generalizations to
multidimensional expansions are also under development \cite{LITTLE}.

Previously the main application of the DVR method was for obtaining bound
state energies and wave functions. For this purpose the wave function is
expanded into a set of $N$ basis functions, whose expansion coefficients are
to be determined. The calculations are of the Galerkin type, namely, the
Hamiltonian applied to the wave function is multiplied on the left by each one
of the expansion basis functions, and the result is integrated over the full
range of the domain of the variable, leading to a set of $N$ linear equations
for the expansion coefficients. The integrals to be evaluated are then
approximated by discrete sums over the values of the integrand evaluated at
the support points times certain weight factors such as in the Gauss
quadrature methods \cite{NIST}.

In the case of the solution of scattering problems the finite element method
(FEM) \cite{FE} has also been developed. In this procedure the radial range is
divided into partitions, also called elements, and the solution of the wave
equation in each partition is expanded into basis functions such as
\textquotedblleft hat\textquotedblright \ functions, Gaussians, or polynomials
of a given order, whose expansion coefficients are to be determined. The
equations for the expansion coefficients are obtained through a Galerkin
procedure, and in many cases the integrals over the basis functions can be
done analytically. The continuity of the wave function from one partition to
the next is achieved by imposing conditions on the expansion coefficients, as
is done for example in Ref. \cite{SB}. In the more recent DVR methods the
basis functions are Lagrange polynomials whose zeros occur at the Lobatto
points \cite{KOPAL}, \cite{KRYLOV}, in which case the quadrature is denoted as
Gauss-Lobatto, and the basis set of functions is denoted as Lagrange-Lobatto.
This basis set was first suggested by Manolopoulos and Wyatt \cite{MAN}, and a
extensive review is given in Ref. \cite{SFB}. The main computational advantage
of using DVR basis functions is that the sum mentioned above reduces to only
one term, because the product of two different DVR functions vanishes at the
support points, and only products between the same DVR functions remain.
Furthermore, within the approximation of the Gauss-Lobatto quadrature rule,
the basis functions are orthogonal. Hence the procedure leads to a discretized
hamiltonian ($N\times N)$ matrix, whose eigenfunctions determine the expansion
coefficients and the eigenvalues determine the bound-state energies. There are
several types of errors introduced by this method. One is due to the
truncation of the expansion of the wave function in terms of basis functions
at an upper limit $N,$ another is due to the approximation of the
Gauss-Lobatto quadrature described above in terms of discrete sums over the
support points. A third error is the accumulation of machine round-off errors.
These errors have been examined for bound state energy eigenvalues,
\cite{WEI}, \cite{BASEL}, \cite{LITTLE}, \cite{SCHN}\ and it is found that the
convergence of the energy with the number $N$ of DVR basis functions is
exponential, and the non-orthogonality error becomes small as $N$ increases.

Very recently a combination of the FE and DVR methods has been introduced into
atomic physics by Rescigno and McCurdy \cite{MCC} for quantum scattering
calculations . These calculations use the FEM approach but in each partition
the basis functions are Lagrange polynomials, and the support points are
Gauss- Lobatto. This \textquotedblleft hybrid\textquotedblright \ method,
denoted as FE-DVR, is now extensively used for atomic physics calculations,
such as for multi electron density distributions in atoms \cite{BBB}, for
photo-ionizing cross sections with fast photon pulses \cite{LINHO},
\cite{HCS}, and for atom-atom scattering calculations \cite{TMR}, to name a
few. However, in these works the accuracy of the results was not studied in
detail. The FE-DVR method is also used extensively for fluid dynamic
calculations since the 1980's \cite{PAT} and also in Seismology \cite{TROMP},
where it is called spectral element method.

The main purpose of the present study is to investigate the accuracy of the
FE-DVR method for the scattering conditions, since all the errors described
above (the Gauss- Lobatto's integration error, the truncation errors of the
expansions, and the accumulation of round-off errors) are still present. In
our study a method of imposing the continuity of the wave function and of the
derivative from one partition to the next is explicitly given, and the
accuracy is obtained by comparing the results of the FE-DVR calculation for
particular solutions of a one dimensional Schr\"{o}dinger equation with a
spectral \cite{SPECTRAL} Chebyshev expansion method \cite{IEM}, S-IEM. The
accuracy of the latter is of the order of $1:10^{-11},$ as is demonstrated in
Appendix A. In our present formulation of the FE-DVR the so-called bridge
functions used in references \cite{MCC}-\cite{TMR} in order to assure the
continuity of the wave function are not used, but are replaced by another method.

In section II the FE-DVR method is described, in section III the accuracy is
investigated by means of numerical examples, and section IV contains the
summary and conclusions. Appendix $A$ contains a short review of the S-IEM
method, in appendix $B$ an estimate of the accumulation of errors is
presented, and in Appendix $C$ some accuracy properties of the finite
difference Numerov (or Milne's) method are presented.

\section{The \ FE-DVR method}

The FE-DVR version of the finite element method differs from the conventional
FEM in that the basis functions for the expansion of the solution $\psi(x)$ in
each partition are\ $N$ \textquotedblleft discrete variable
representation\textquotedblright \ (DVR) functions, which in the present case
are Lagrange polynomials $\ell_{i}(x)$, $i=1,2,..N$ of a given order $N-1,$
\begin{equation}
\ell_{i}(x)=%
{\displaystyle \prod_{k=1}^{N}}
\frac{(x-x_{k})}{(x_{i}-x_{k})},~~k\neq i \label{1g}%
\end{equation}
defined for example in Eq. $(25.2.2)$ of Ref. \cite{AS}, and in section 3.3(i)
of Ref. \cite{NIST}. These functions are widely used for interpolation
procedures and are described in standard computational textbooks. This FE and
DVR combination was introduced in Ref. \cite{MCC}, and has the advantage that
integrals involving these polynomials amount to sums over the functions
evaluated only at the support points. In the present case the support points
are Lobatto points $x_{j}$ and weights $w_{j}$, $j=1,2,..N,$ defined in Eq.
$(25.4.32)$ of Ref. \cite{AS}, in terms of which a quadrature over a function
$f(x)$ in the interval $[-1,+1]$ is approximated by%
\begin{equation}
\int_{-1}^{+1}f\left(  x\right)  dx\simeq \sum_{j=1}^{N}f(x_{j})w_{j}.
\label{2a}%
\end{equation}
If $f$ is a polynomial of degree $\leq2N-3$ then Eq. (\ref{2a}) will be exact.
This however is not the case for the product of two Lagrange polynomials
$\ell_{i}(x)\ell_{j}(x),$ a polynomial of order $2(N-1).$ In the Gauss-Lobatto
quadrature approximation \cite{KOPAL}, \cite{KRYLOV}, given by the right hand
side of Eq. (\ref{2a}), these Lagrange polynomials are orthogonal to each
other, but they are not rigorously orthogonal \cite{BASEL} because the left
hand side of Eq. (\ref{2a}) is not equal to the right hand side. If the
integral limits are different from $\pm1$, such as $\int_{a}^{b}f(r)dr$
then\ the variable $r$ can be scaled to the variable $x.$ Our method differs
from that of Ref. \cite{MCC} in that we do not use their \textquotedblleft
bridge\textquotedblright \ functions, but rather insure continuity of the
solution and its derivative from one partition to the next by using only the
Lagrange functions. Since the Lobatto points are not evenly spaced, expansion
(\ref{2a}) converges uniformly, which is a general feature of spectral methods
\cite{SPECTRAL}. A further $DVR$ advantage is that the Gauss-Lobato
approximation of the integral
\begin{equation}
\int_{-1}^{1}\ell_{i}(x)f\left(  x\right)  \ell_{j}(x)dx\simeq \delta
_{i,j}w_{j}f(x_{i}) \label{2b}%
\end{equation}
is diagonal in $i,j$ and is given by only one term. The convolution
\begin{equation}
\int_{-1}^{1}\ell_{i}(x)\int_{-1}^{1}K\left(  x,x^{\prime}\right)  \ell
_{j}(x^{\prime})dx^{\prime}\ dx\simeq w_{i}w_{j}K(x_{i},x_{j}) \label{2c}%
\end{equation}
is also approximated by one non-diagonal term only, which is a marked
advantage for solving nonlocal or coupled channel Schr\"{o}dinger equations.
The kinetic energy integral can be expressed in the form
\begin{equation}
\int_{-1}^{1}\ell_{i}(x)\frac{d^{2}}{dx^{2}}\ell_{j}(x)dx=-\int_{-1}^{1}%
\ell_{i}^{\prime}(x)\ell_{j}^{\prime}(x)dx+\delta_{i,N}\ell_{j}^{\prime
}(1)-\delta_{i,1}\ell_{j}^{\prime}(-1) \label{2d}%
\end{equation}
after an integration by parts. In the above the prime denotes $d/dx.$ The
integral on the right hand side of this equation can be done exactly with the
Gauss-Lobatto quadrature rule (\ref{2a}), since the integrand is a polynomial
of order $2N-4$, that is less than the required $2N-3.$

For the case of a local potential $V$ with angular momentum number $L=0$ the
equation to be solved is%
\begin{equation}
\left(  \frac{d^{2}}{dr^{2}}+k^{2}\right)  \psi(r)=V(r)\psi(r), \label{11}%
\end{equation}
and for a nonlocal potential $K$, the term $V(r)\psi(r)$ is replaced by
$\int_{0}^{\infty}K(r,r^{\prime})\psi(r^{\prime})dr^{\prime}.$ The wave number
$k$ is in units of $fm^{-1}$ and the potential $V$ is in units of $fm^{-2}$,
where quantities in energy units are transformed to inverse length units by
multiplication by the well known factor $2m/\hslash^{2}.$ In the scattering
case the solutions $\psi(r)$ are normalized such that for $r\rightarrow \infty$
they approach
\begin{equation}
\psi(r)\rightarrow \sin(kr)+\tan(\delta)\cos(kr), \label{B9}%
\end{equation}
and with that normalization one finds
\begin{equation}
\tan(\delta)=-\frac{1}{k}\int_{0}^{\infty}\sin(kr)\ V(r)\  \psi(r)dr,
\label{B10}%
\end{equation}
as is well known \cite{LANDAU}.

The FE-DVR procedure is as follows. We divide the radial interval into $N_{J}$
partitions (also called elements in the finite element calculations
\cite{FE}), and in each partition we expand the wave function into $N$
Lagrange functions $\ell_{i}(r),i=1,2,..N$%
\begin{equation}
\psi^{(J)}(r)=\sum_{i=1}^{N}c_{i}^{(J)}\ell_{i}(r),~~\ b_{1}^{(J)}\leq
r\leq \bigskip b_{2}^{(J)}, \label{1a}%
\end{equation}
The starting and end points of each partition are denoted as $b_{1}^{(J)}$ and
$b_{2}^{(J)}$, respectively. We define the value and the derivative of the
wave function at the end point of the previous partition as
\begin{equation}
\psi^{(J-1)}(b_{2}^{(J-1)})=c_{N}^{(J-1)}, \label{1b}%
\end{equation}
where $c_{N}^{(J-1)}$ is the last coefficient of the expansion (\ref{1a}) of
$\psi^{(J-1)}$, and
\begin{equation}
A^{(J-1)}\equiv \frac{d}{dr}\psi^{(J-1)}(b_{2}^{(J-1)})=\sum_{i=1}^{N}%
c_{i}^{(J-1)}\ell \ _{i}^{\prime}(b_{2}^{(J-1)}), \label{1c}%
\end{equation}
respectively, where $\ell \ _{i}^{\prime}(r)=d\ell_{i}(r)/dr.$ The result
(\ref{1b}) follows from the fact that that $\ell_{i}(b_{2})=0$ for
$i=1,2,..N-1$, and $\ell_{N}(b_{2})=1$.\ For the first partition we
arbitrarily take a guessed value of $A^{(0)}$ for the non-existing previous
partition$,$ and later renormalized the whole wave function by comparing it to
a known value. That is equivalent to renormalizing the value of $A^{(0)}$. In
finite element calculations continuity conditions of the wave function from
one partition to the next are also imposed. However, the method described
below applies specifically to the case that the basis functions in each
element are of the DVR type, rather than general polynomials of a given order
\cite{SB}.

By performing the Galerkin integrals of the Schr\"{o}dinger Eq. over the
$\ell_{i}$ in each partition $J$
\begin{align}
\langle \ell_{i}(T+V-k^{2})\psi^{(J)}\rangle &  =\label{1d}\\
\int_{b_{1}^{(J)}}^{b_{2}^{(J)}}\ell_{i}(r)(T+V-k^{2})\psi^{(J)}(r)dr  &
=0,~~\ i=1,2,..N
\end{align}
we obtain a homogeneous matrix equation in each partition for the coefficients
$c_{i}^{(J)},$ $i=1,2,..N$\
\begin{equation}
M^{(J)}\  \vec{c}^{(J)}=0, \label{1e}%
\end{equation}
where $\vec{c}^{(J)}$ represents the ($N\times1)$ column vector of the
coefficients $c_{i}^{(J)},$\ and where the matrix elements of $M$ are given by
$M_{ij}=\langle \ell_{i}(T+V-k^{2})\ell_{j}\rangle.$ Here $T=-d^{2}/dr^{2}.$
The continuity conditions are imposed by transforming the homogeneous equation
(\ref{1e}) of dimension $N$ into an inhomogeneous equation of dimension $N-2$
whose driving terms are composed of the function $\psi$ and $d\psi/dr$
evaluated at the end of the previous partition. These conditions are given by
\begin{equation}
c_{1}^{(J)}=c_{N}^{(J-1)} \label{1f}%
\end{equation}
where use has been made of $\ell_{i}(b_{1})=0$ for $i=2,..N$, and $\ell
_{1}(b_{1})=1,$ and%
\begin{equation}
\frac{d\  \psi^{(J-1)}(b_{2}^{(J-1)})}{dr}=\sum_{i=1}^{N}c_{i}^{(J)}\ell
\ _{i}^{\prime}(b_{1}^{(J)})=A^{(J-1)}. \label{1j}%
\end{equation}
These two conditions can be written in the matrix form
\begin{equation}
F_{11}\alpha+F_{12}\beta=\gamma \label{2}%
\end{equation}
where
\begin{equation}
F_{11}=%
\begin{pmatrix}
1 & 0\\
\ell_{1}^{\prime} & \ell_{2}^{\prime}%
\end{pmatrix}
_{b_{1}^{(J)}}^{(J)};~~F_{12}=%
\begin{pmatrix}
0 & 0 & \cdots & 0\\
\ell_{3}^{\prime} & \ell_{4}^{\prime} & \cdots & \ell_{N}^{\prime}%
\end{pmatrix}
_{b_{1}^{(J)}}^{(J)}, \label{3}%
\end{equation}
where
\begin{equation}
\alpha=%
\begin{pmatrix}
c_{1}\\
c_{2}%
\end{pmatrix}
^{(J)}, \label{3a}%
\end{equation}
where
\begin{equation}
\beta=%
\begin{pmatrix}
c_{3}\\
c_{4}\\
\vdots \\
c_{N}%
\end{pmatrix}
^{(J)}, \label{3b}%
\end{equation}
and where
\begin{equation}
\gamma=%
\begin{pmatrix}
c_{N}\\
A
\end{pmatrix}
^{(J-1)}. \label{3c}%
\end{equation}

With that notation Eq. (\ref{1e}) can be written in the form%
\begin{equation}%
\begin{pmatrix}
M_{11} & M_{12}\\
M_{21} & M_{22}%
\end{pmatrix}%
\begin{pmatrix}
\alpha \\
\beta
\end{pmatrix}
=0, \label{1}%
\end{equation}
where the matrix $M^{(J)}$ has been decomposed into four submatrices
$M_{11},M_{12},M_{21}$, and $M_{22},$ which are of dimension $2\times
2,2\times(N-2),(N-2)\times2,$ and $(N-2)\times(N-2),$ respectively. The column
vector $\alpha$ can be eliminated in terms of $\beta$ and $\gamma$ by using
Eq. (\ref{2}),
\begin{equation}
\alpha=F_{11}^{-1}(-F_{12}\beta+\gamma) \label{6}%
\end{equation}
and the result when introduced into Eq. (\ref{1}) leads to an inhomogeneous
equation for $\beta$%
\begin{equation}
(-M_{21}F_{11}^{-1}F_{12}+M_{22})\beta=-M_{21}F_{11}^{-1}\gamma. \label{5}%
\end{equation}
Once the vector $\beta$ is found from Eq. (\ref{5}), then the components of
the vector $\alpha$ can be found from Eq. (\ref{6}), and the calculation can
proceed to the next partition.

If one expresses the inverse of $F_{11}$ analytically%
\begin{equation}
F_{11}^{-1}=%
\begin{pmatrix}
1 & 0\\
-\  \frac{\ell_{1}^{\prime}}{\ell_{2}^{\prime}} & \frac{1}{\ell_{2}^{\prime}}%
\end{pmatrix}
. \label{7}%
\end{equation}
then one finds
\begin{equation}
F_{11}^{-1}\gamma=%
\begin{pmatrix}
c_{N}^{(J-1)}\\
-\  \frac{\ell_{1}^{\prime}}{\ell_{2}^{\prime}}c_{N}^{(J-1)}+\frac{A^{(J-1)}%
}{\ell_{2}^{\prime}}%
\end{pmatrix}
\label{8}%
\end{equation}
and%
\begin{equation}
F_{11}^{-1}F_{12}=%
\begin{pmatrix}
0 & 0 & \cdots & 0\\
\frac{\ell_{3}^{\prime}}{\ell_{2}^{\prime}} & \frac{\ell_{4}^{\prime}}%
{\ell_{2}^{\prime}} & \cdots & \frac{\ell_{N}^{\prime}}{\ell_{2}^{\prime}}%
\end{pmatrix}
. \label{9}%
\end{equation}
Inserting (\ref{7}) into (\ref{6}) one finds that $c_{1}^{(J)}=c_{N}^{(J-1)}$,
but $c_{2}^{(J)}$ is a function of $c_{N}^{(J-1)}$, $A^{(J-1)},$ and the
vector $\beta.$

\section{Accuracy}

\begin{figure}
[ptb]
\begin{center}
\includegraphics[
height=2.3903in,
width=3.1799in
]%
{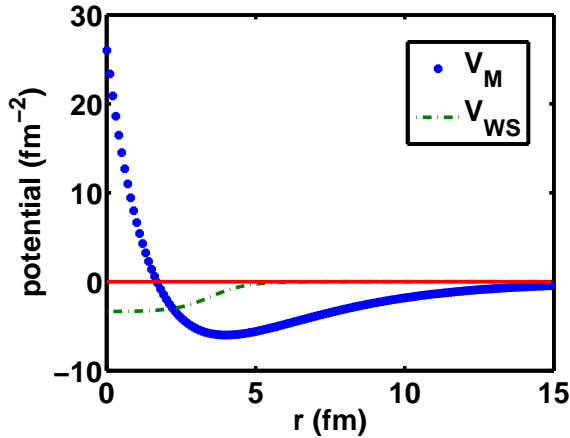}%
\caption{ (Color on line) Potentials Morse $V_{M}$ \ and Woods Saxon $V_{WS}$
as a function of radial distance $r$. These potentials are given by Eqs.
(\ref{B7}) and (\ref{B8}), respectively.}%
\label{FIG4}%
\end{center}
\end{figure}

\begin{figure}
[ptb]
\begin{center}
\includegraphics[
height=2.1456in,
width=2.853in
]%
{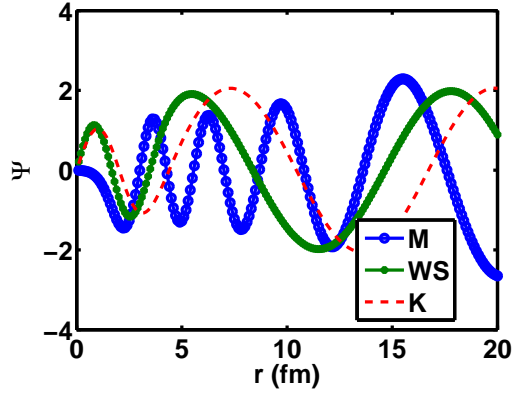}%
\caption{(Color on line) The wave functions for the local potentials $V_{M}$
and $V_{WS},$ and for the nonlocal potential $K$, described in the text. The
wave number is $k=0.5$ $fm^{-1}$ and the potentials $V_{M}$ and $V_{WS}$ are
illustrated in Fig. \ref{FIG4}.}%
\label{FIG5}%
\end{center}
\end{figure}

We have tested the accuracy for cases with angular momentum $L=0$ for two
local potentials $V_{M}$ and $V_{WS},$ shown in Fig. \ref{FIG4}, and for a
nonlocal potential $K(r,r^{\prime})$ of the Perey-Buck type \cite{PB}.
Potential $V_{M}$ is of a Morse type with a repulsive core near the origin,
given by
\begin{equation}
V_{M}(r)=6\exp(-0.3\ r+1.2)\times \left[  \exp(-0.3\ r+1.2)-2\right]  .
\label{B7}%
\end{equation}
and $V_{WS}$ \ is a short-ranged simple Woods-Saxon potential given by
\begin{equation}
V_{WS}(r)=-3.36/\{1+\exp[(r-3.5)/0.6]\}. \label{B8}%
\end{equation}
The coefficients $6$ and $-3.36$ are in units of $fm^{-2}$, the distances $r$
are in units of $fm$, and all other factors are such that the arguments of the
exponents are dimensionless. These potentials are shown in Fig. \ref{FIG4},
and the respective wave functions are shown in Fig. \ref{FIG5}.
\ The choice of these potentials is motivated by the difference in the degree
of computational difficulty that they offer in the solution of the
Schr\"{o}dinger equation. Potential $V_{WS}$ has no repulsive core near the
origin and is of short range. Hence the corresponding wave function does not
have large derivatives near the origin, and needs not to be calculated out to
distances larger than $20$\ $fm$, where the potential is already negligible,
of the order or $10^{-11}$. By contrast, neither of these two features apply
for the case of $V_{M}$. In order to obtain an accuracy of $1:10^{-11}$ the
wave function has to be calculated out to $100\ fm$, as is indeed done in the
calculation of the bench mark S-IEM solution, and the repulsive core near the
origin is more difficult to treat. The nonlocal potential $K$ is described in
Eq. ($3$) of Ref. \cite{RPB} together with the Appendix of Ref. \cite{PB}. The
accuracy of the corresponding wave function obtained with the S-IEM method for
this nonlocal potential is illustrated in Fig. $7$ of Ref. \cite{RPB}. For the
nonlocal case only one partition is used in the FE-DVR method, that extends
from $r=0$ to $R_{\max},$ but in view of Eq. (\ref{2c}), the calculation is
very efficient.

In order to ascertain the accuracy of the FE-DVR method, the solutions of Eq.
(\ref{11}) are compared with the solutions obtained by the spectral integral
equation method (S-IEM) \cite{IEM}, whose accuracy is $1:10^{-11},$ as
described in Appendix A. The numerical FE-DVR solutions are first normalized
by comparison with the S-IEM solutions at one chosen radial position near the
origin, and the error of the normalized FE-DVR function is determined by
comparison with the S-IEM function at all other radial points $r.$ Since the
S-IEM function depends on the values of the potential at all points $[0\leq
r\leq R_{\max}],$ the S-IEM\ calculation has to be carried out to a distance
$R_{\max}$ large enough so that the contribution from $V(r\geq R_{\max})$ is
smaller than the desired accuracy of the S-IEM\  \ solution. The same is not
the case for the FE-DVR solutions $\psi_{FE-DVR}(r),$ since the un-normalized
solution depends only on the potentials for distances less than $r$. However,
if the normalization of the wave function (\ref{B9}) is to be accomplished by
matching it to $\sin(kr)$ and $\cos(kr)$ at $R_{\max}$ in the asymptotic
region, then the numerical errors that accumulated out to $R_{\max}$ will
affect the wave function at all distances. These errors can be avoided by an
iterative procedure for the large distance part of the wave function, as will
be described in a future publication \cite{RAW-ITER}.

\begin{figure}
[ptb]
\begin{center}
\includegraphics[
height=2.4068in,
width=3.1955in
]%
{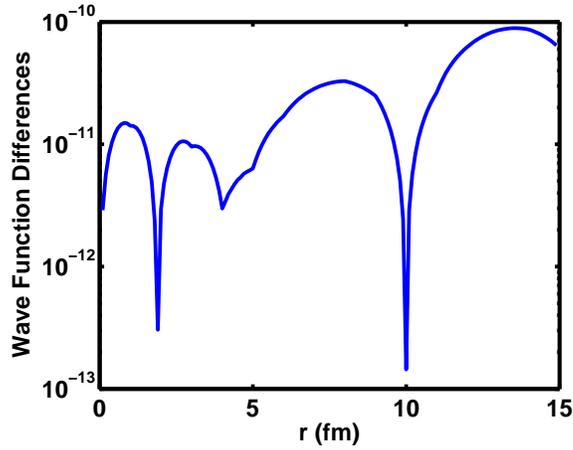}%
\caption{(Color on line) Accuracy of the FE-DVR solution of the
Schr\"{o}dinger Eq. for the Woods Saxon potential $V_{WS}$ displayed in Fig.
\ref{FIG4}. The wave number is $k=0.5\ fm^{-1}$, the size of each partition is
$1.fm$, and there are $20\ $Lobatto points per partition. The graph shows the
accuracy of the wave function $\psi$ by displaying the absolute value of the
difference between the FE-DVR and the S-IEM wave functions. The latter is
deemed accurate to $1:10^{-11}$.}%
\label{FIG2}%
\end{center}
\end{figure}

\begin{figure}
[ptb]
\begin{center}
\includegraphics[
height=2.3531in,
width=3.122in
]%
{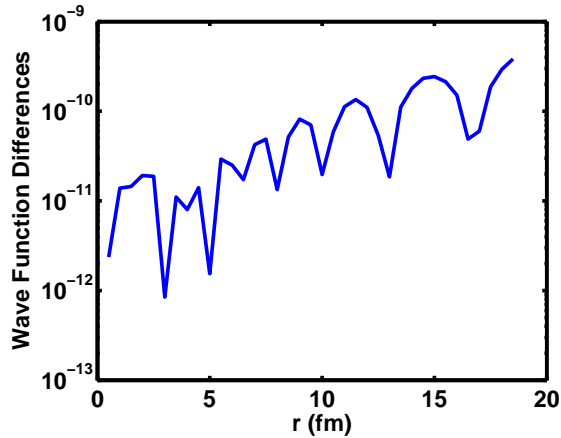}%
\caption{(Color on line) The accuracy of the FE-DVR\ wave function for the
potential $V_{M}$ as obtained by comparison with the S-IEM result. The latter
is accurate to $1:10^{-11}.$ The wave number is $k=0.5\ fm^{-1},$ the number
of Lobatto points per partition is $20$, the size of each partition is
$1\ fm.$}%
\label{FIG3}%
\end{center}
\end{figure}

\begin{figure}
[ptb]
\begin{center}
\includegraphics[
height=2.207in,
width=2.9283in
]%
{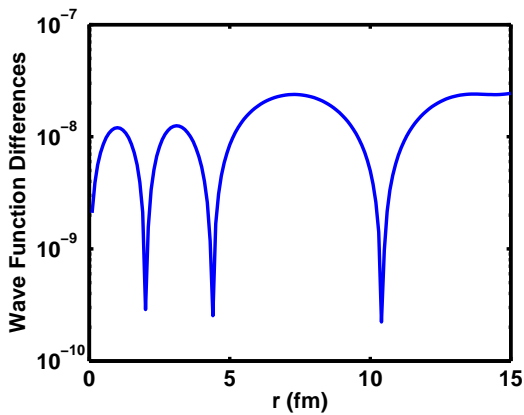}%
\caption{ (Color on line) Same as\ Fig. \ref{FIG2} for the nonlocal Perey-Buck
potential $K(r,r\prime)$. The wave number is $k=0.5fm^{-1}$, only one
partition was used in the full radial interval from $0$ to $15$ $fm$ using a
total of $130$ Lobatto grid points. The accuracy of $10^{-8}$ is consistent
wth the estimate made in Eq. (\ref{A4}) in Appendix B.}%
\label{FIG6}%
\end{center}
\end{figure}

The results for potentials $V_{WS}$ and $V_{M}$ are shown figures \ref{FIG2}
and \ref{FIG3}, respectively. In both cases the error of the wave function
starts with $10^{-11}$ at the small distances, and increases to $10^{-10}$ as
the distance increases, due to the accumulation of various errors. The
accuracy for the nonlocal potential $K$ is shown in Fig. \ref{FIG6}. The
accuracy of the integral (\ref{B10}), for a fixed size of all partitions as a
function of the number $N$ of Lobatto points in each partition, is shown in
Fig. \ref{FIG14}, where the open circles represent an upper limit of the
estimated accuracy as developed in Appendix B, of order $(N-2)^{3}.$ This
figure is important because it shows the nearly exponential increase of
accuracy as $N$ increases, until the accumulation of errors overwhelms this
effect once the value of $N$ increases beyond a certain value, $20$ for the
case of Fig. \ref{FIG14}. The accuracy of the integral (\ref{B10}) for a fixed
number $N$ per partition, but for several different partition sizes, is
displayed in Fig. \ref{FIG15}. This figure shows that the accuracy decreases
exponentially with the size of the partition, which can be interpreted as an
exponential increase of the accuracy with the number of Lobatto points in each
partition of fixed length.

Finally, the FE-DVR computing time as a function of the number $N$ of Lobatto
points in each partition is displayed in Fig. \ref{FIG16}, where it is also
compared with an estimate described in Appendix $B$ of the number of floating
point operations expected. According to this estimate, the time per floating
point operation turns out to be $\simeq10^{-8}$ in a MATLAB computation
performed on a desktop using an Intel TM2 Quad, with a CPU Q 9950, a frequency
of 2.83 GHz, and a RAM\ of 8 GB. The dashed line represents the total time
required for a comparable S-IEM computation. That comparison shows that the
FE-DVR method can be substantially faster than the S-IEM even though the
former has many more support points, depending on the radial range and on the
accuracy required. Further details are given in Table \ref{TABLE2} in Appendix A.

\bigskip%

%


%

\bigskip \medskip%

\begin{figure}
[ptb]
\begin{center}
\includegraphics[
height=2.4344in,
width=3.2353in
]%
{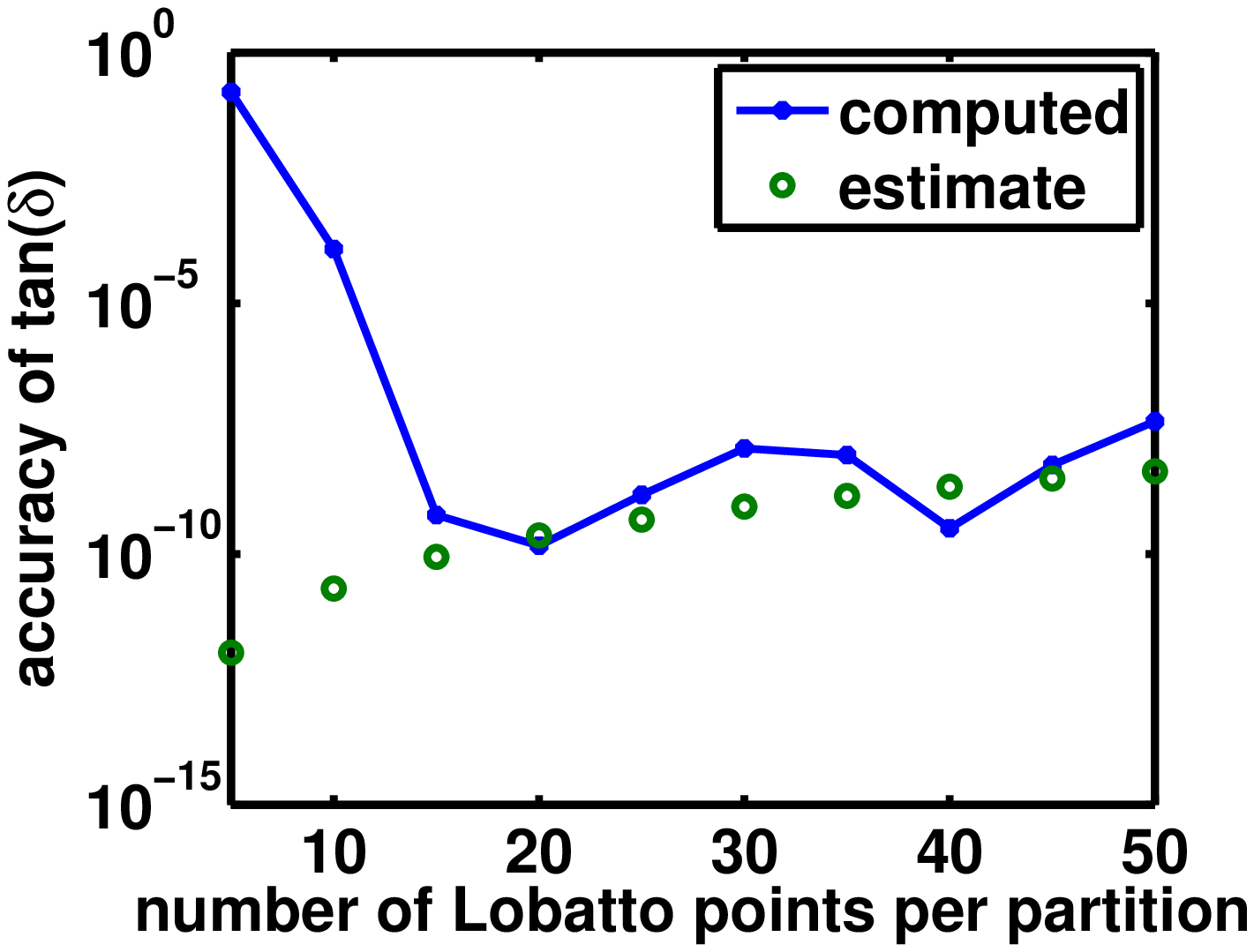}%
\caption{(Color on line) Accuracy of the integral $\int_{0}^{100}%
\sin(kr)\ V_{M}(r)\  \psi(r)\ dr,$ obtained with the FE-DVR method as a
function of the number of Lobatto points in each partition. The length of each
partition is $1.0\ fm$, the number of partitions is $100$. The potential is
$V_{M}$, the wave number is $k=0.5\ fm^{-1}$, the accuracy is obtained by
comparisom with the S-IEM result which is accurate to $1:10^{-11}$. The open
circles represent an estimate of the upper bound for the accumulation of
roundoff errors, given by Eq. (\ref{A1}) in Appendix B.}%
\label{FIG14}%
\end{center}
\end{figure}

\begin{figure}
[ptb]
\begin{center}
\includegraphics[
height=2.073in,
width=2.7562in
]%
{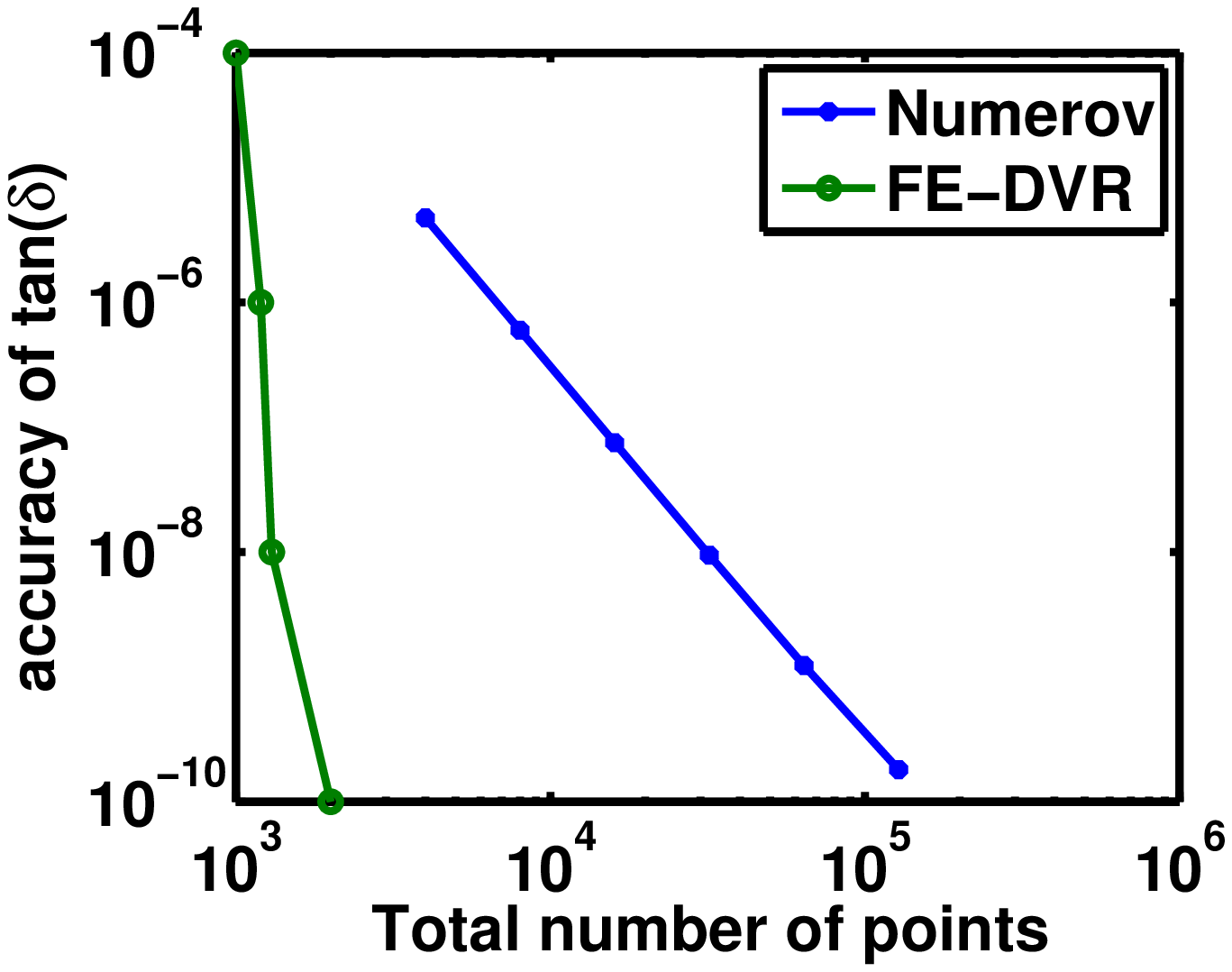}%
\caption{(Color on line) This accuracy comparison\ for $\tan(\delta)$ is
performed for the potential $V_{M}$ and $k=0.5\ fm^{-1}$ in the radial
interval $[0,100fm]$. The partition sizes in the FE-DVR method have a length
of $1\ fm$ each, and the number of Lobatto points in each partition is given
by $1/100\ th$ of the total number of points. Numerov is a 6th order finite
difference method with equidistant points, as described in Appendix C.}%
\label{FIG24}%
\end{center}
\end{figure}

A comparison between the FE-DVR and a finite difference sixth order Numerov
method \ of the accuracy of $\tan(\delta)$ is illustrated in Fig.
\ref{FIG24}.
This comparison shows that for an accuracy of $\tan(\delta)$ of $\simeq
10^{-8}$, the FE-DVR method requires $15$ times fewer meshpoints, and is
approximately $100$ times faster than the Numerov method. More details are
presented in Appendix C.
\begin{figure}
[ptb]
\begin{center}
\includegraphics[
height=2.271in,
width=3.0199in
]%
{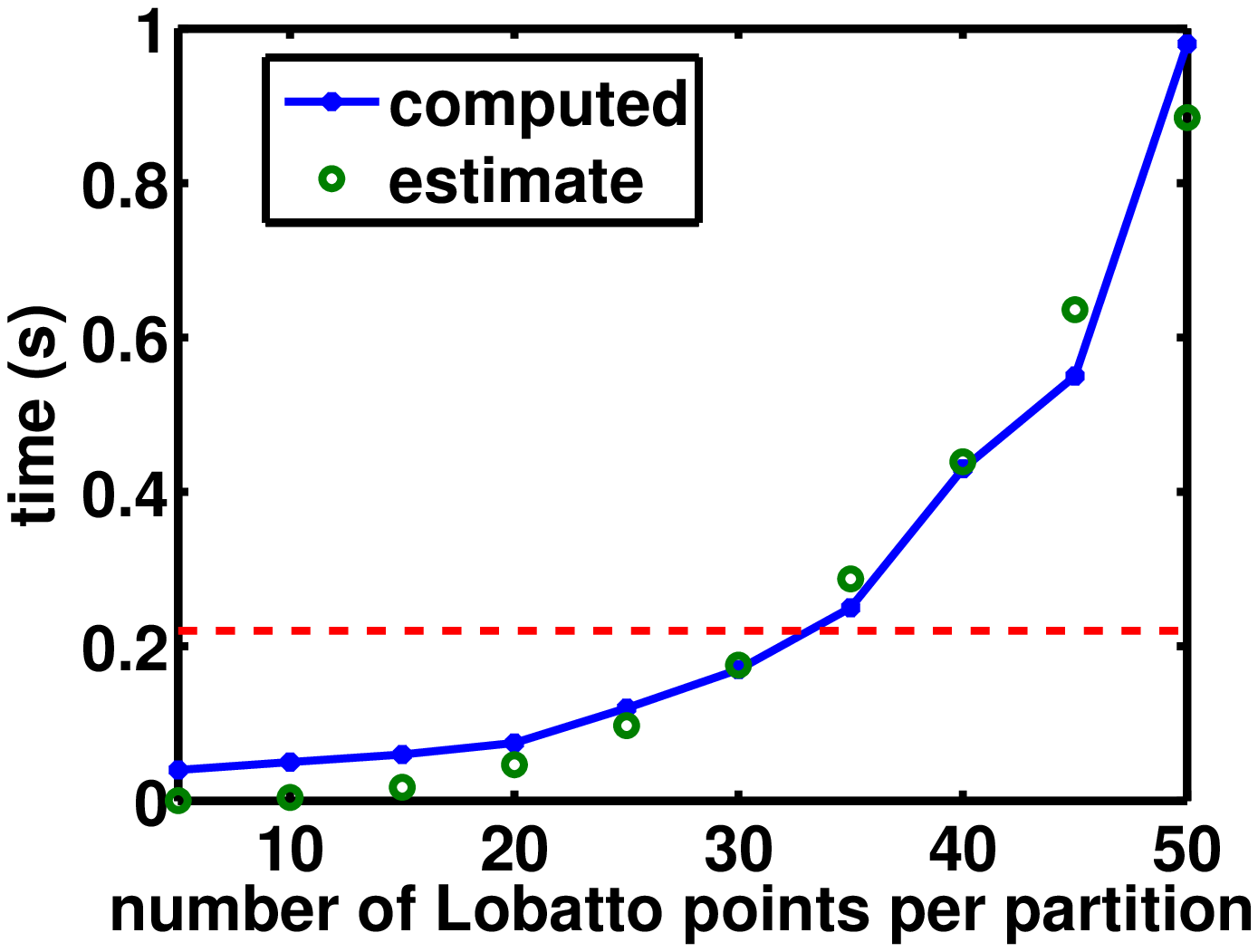}%
\caption{(Color on line) The computing time in MATLAB for the calculations
described in Fig. \ref{FIG14}. The estimate is given by Eq. (\ref{A1}), with
the factor $10^{-16}$ replaced by $2\times10^{-8}.$ The latter represents the
time for each floating point operation. The dashed line represents the
computing time for the S\_IEM calculation, described in Fig. \ref{FIG14}.}%
\label{FIG16}%
\end{center}
\end{figure}
\begin{figure}
[ptb]
\begin{center}
\includegraphics[
height=2.2399in,
width=2.9784in
]%
{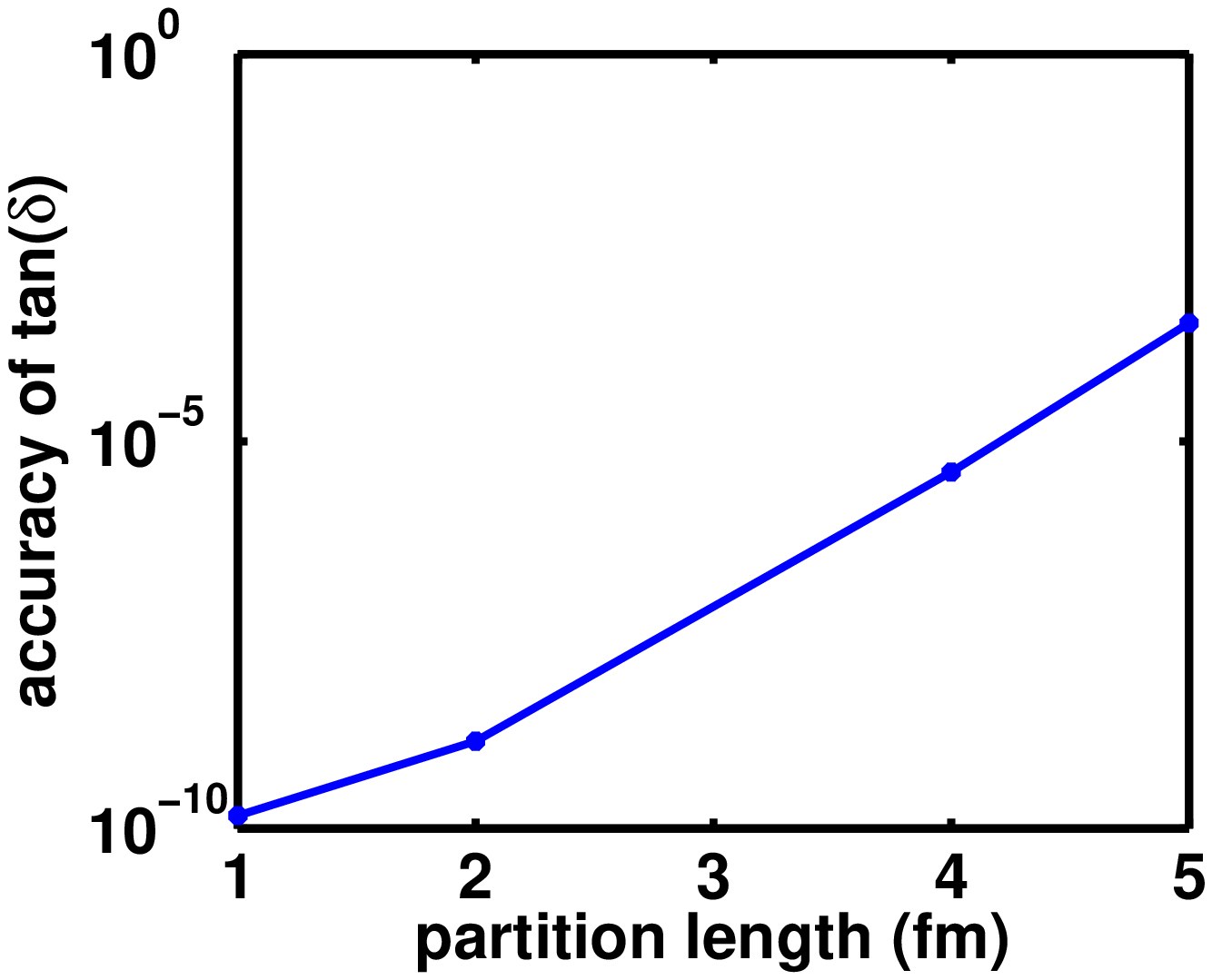}%
\caption{(Color on line) Accuracy of the integral $\int_{0}^{100}%
\sin(kr)\ V_{M}(r)\  \psi(r)\ dr$, Eq. (\ref{B10}), as a function of the length
of each partition, into which the radial interval $[0,100\ fm]$ is divided.
The total number $N=20$ of Lobatto points in each partition is kept constant.
The conditions are the same as in Fig. \ref{FIG14}. This figure shows that the
accuracy decreases exponentially with the size of the partition. For $20$
partitions the computation time is $0.060$ s, for $100$ partitions it is
$0.075$ s. \ }%
\label{FIG15}%
\end{center}
\end{figure}

\section{Summary and Conclusions}

The accuracy of a hybrid finite element method (FE-DVR) has been examined for
the solution of the one dimensional Schr\"{o}dinger equation with scattering
boundary conditions. This method\  \cite{MCC} uses as basis functions the
discrete variable representation Lagrange polynomials $\ell_{i}(r)$,
$i=0,1,2,..N,$ on a mesh of $N$ Lobatto support points. The accuracy of the
FE-DVR method is obtained by comparison with a spectral method S-IEM, whose
accuracy is of the order of $1:10^{-11}$. An important advantage of a discrete
variable representation basis is the ease and accuracy with which integrals
can be performed using a Gauss-Lobatto integration algorithm that furthermore
render the matrix elements $\langle \ell_{i}(V-E)\ell_{j}\rangle$ diagonal.
This feature permits one to easily solve the Schr\"{o}dinger Eq. also\ in the
presence of nonlocal potentials with a kernel of the form $K(r,r^{\prime}),$
as is demonstrated in one of our numerical examples. Another advantage is that
the Galerkin matrix elements of the kinetic energy operator $T$ need not be
recalculated anew for each partition because they are the same in all
partitions to within a normalization factor that only depends on the size of
the partition. A further advantage is that the convergence of the expansion
(\ref{1a}) with the number $N$ of basis functions is exponential, in agreement
of what it is the case for bound state finite element calculations with
Lobatto discretizations \cite{BASEL}. A possible disadvantage may be that if
the number of the Lagrange polynomials in each partition is very large and/or
the number of partitions is large, as is the case for long ranged potentials,
then the accumulation of roundoff and algorithm errors may become unacceptably large.

In summary, for scattering solutions of the Schr\"{o}dinger equation the
accuracy of the FE-DVR method increases exponentially with the number of
Lagrange polynomials in each partition until the accumulation of roundoff and
truncation errors overwhelm the result. The FE-DVR can easily achieve an
accuracy of the order of $10^{-10}$ for the scattering phase shifts for either
local or nonlocal short ranged potentials; it is less complex than the
spectral S-IEM method but is comparable in the amount of computing time; and,
in addition, it is substantially more efficient than a finite difference
Numerov method. The latter result is demonstrated by the fact that the FE-DVR
was found to be a hundred times faster than the Numerov for an accuracy of
$10^{-8}$ of the scattering phase shift.

Acknowledgements:\ One of the authors (GR) is grateful to Professor McCurdy
for a stimulating conversation on the use of Lagrange polynomials in finite
element calculations.\bigskip \bigskip

{\LARGE Appendix A: the S-IEM method}

A version of the spectral method employed here was developed recently
\cite{IEM}. It consists in dividing the radial interval into partitions of
variable size, and obtaining two independent solutions of the Schr\"{o}dinger
Eq. (\ref{11}) in each partition , denoted as $Y(x)$ and $Z(x)$. These
solutions are obtained by transforming Eq. (\ref{11}) into an equivalent
Lippmann-Schwinger integral equation (L-S) and solving the latter by expanding
the solution into Chebyshev functions, mapped to the interval $[-1,+1]$. The
corresponding discretized matrices are not sparse, but are of small dimension
equal to the number of Chebyshev points per partition. The solution $\psi$ in
each partition is obtained by a linear combination of the two independent
functions $Y(x)$ and $Z(x)$, with coefficients that are determined from the
solution of a matrix equation of dimension twice as large as the number of
partitions, but the corresponding matrix is sparse. Details are given in Ref.
\cite{IEM}, and a pedagogical version is found in Ref. \cite{STRING}.

\begin{figure}
[ptb]
\begin{center}
\includegraphics[
height=2.2139in,
width=2.9438in
]%
{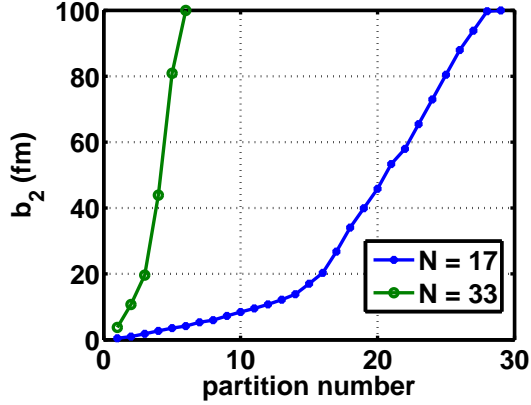}%
\caption{(Color on line) The partition distrtibution for the S-IEM method in
the radial interval $[0,100\ fm]$. for two different numbers $N$ of the
Chebyshev expansion functions in each partition. The end point $b_{2}$ of each
partition is shown on the vertical axis, and the corresponding partition
number is shown on the horizontal axis. The potential is $V_{M}$ , and the
wave number is $k=0.5\ fm^{-1}$. The accuracy parameter $tol$ in each
partition is $10^{-12}.$ The computation time for each case is approximately
the same, $0.2\ s,$and the accuracy of the wave function in both cases is the
same, $1:10^{-11}.$}%
\label{FIG5A}%
\end{center}
\end{figure}

One of the features of the S-IEM method is that the size of each partition is
adaptively determined such that the accuracy of the functions $Y(x)$ and
$Z(x)$ is equal or better than a pre-determined accuracy parameter $tol$,
which in the present case is $tol=10^{-12}.$ In the region where the potential
$V$ is small the corresponding partition size is large. When the number of
Chebyshev expansion functions $N$ per partition is large the size of the
partitions is correspondingly large. As is illustrated in Fig. \ref{FIG5A}
when $N$\ is increases from $17$ to $33$ the number of partitions decreases
from $29$ to $6$, yet the accuracy of the respective wave functions is
approximately the same, $1:10^{-11},$ and the computing time is also
approximately the same, $0.2s.$%

\begin{figure}
[ptb]
\begin{center}
\includegraphics[
height=2.3237in,
width=3.0891in
]%
{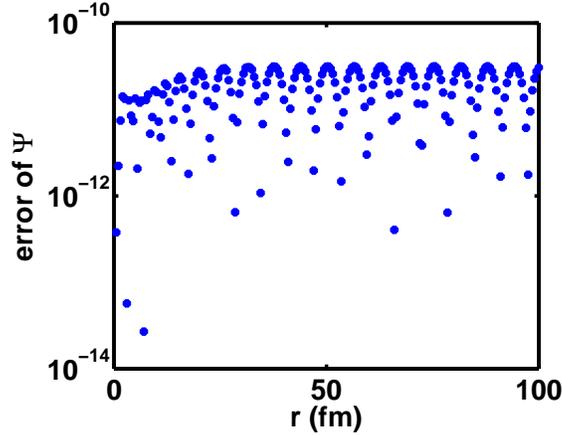}%
\caption{(Color on line) The y-axis illustrates the absolute value of the
difference between two S-IEM wave functions, calculated with accuracy
parameters $tol=10^{-11}$ and $10^{-12},$ respectively, for potential $V_{M}$
and $k=0.5\ fm^{-1}$. This difference is less than $4\times10^{-11}$ for all
values of $r.$}%
\label{FIG13}%
\end{center}
\end{figure}

For the present S-IEM benchmark calculations the value of $N$ is $17$, and for
the case of $V_{M}$ the maximum value of $r$ is $100\ fm.$ Such a large value
is required because the potential decays slowly with distance and becomes less
in magnitude than $5\times10^{-12}$ only beyond $r=100\ fm$. Had the potential
been truncated at a smaller value of $r,$ then the truncation error would have
propagated into all values of the wave function and rendered it less accurate.
The accuracy of the S-IEM wave function can be seen from Fig. \ref{FIG13},
which compares two S-IEM wave functions with accuracy parameters
$tol=10^{-11}$ and $10^{-12}$, respectively$.$ The result is that the accuracy
of the IEM wave function for $N=17$ and $R_{\max}=100fm$ and $tol=10^{-11}$ is
$4\times10^{-11},$ and that for $tol=10^{-12}$ the accuracy is better than
$10^{-11}.$

The wave functions are normalized such that their asymptotic value is given by
Eq. (\ref{B9}). The corresponding values of $\tan(\delta)$, Eq. (\ref{B10}),
for potentials $V_{M}$ and $V_{WS}$ and a wave number $k=0.5\ fm^{-1}$ are
$2.6994702502$ and $-1.7107344227$, respectively. Table \ref{TABLE1} shows the
number of partitions, the accuracy of $\tan(\delta)$, and the computing time
of the S-IEM method for various tolerance parameters inputted into the code
for the potential $V_{M},$ with $k=0.5.$ The number of Chebyshev polynomials
in each partition is $17$, the total number of points displayed in the third
column is equal to $17$ times the number of partitions. The error of
$\tan(\delta)$ is obtained by comparing the value of tan$(\delta)$ for a
particular tolerance parameter with the value obtained for $tol=10^{-12}.$
\begin{table}[tbp] \centering
\begin{tabular}
[c]{|c|c|c|c|c|}\hline
$Tol.$ & $Part^{\prime}ns$ & $Points$ & $Err[$tan$(\delta)]$ & $time\ (s)$%
\\ \hline \hline
$10^{-12}$ & $37$ & $629$ & $-$ & $0.178$\\ \hline
$10^{-10}$ & $25$ & $425$ & $4.6\times10^{-12}$ & $0.181$\\ \hline
$10^{-8}$ & $17$ & $289$ & $7.7\times10^{-11}$ & $0.171$\\ \hline
$10^{-6}$ & $11$ & $187$ & $5.2\times10^{-7}$ & $0.165$\\ \hline
$10^{-4}$ & $7$ & $119$ & $2.8\times10^{-4}$ & $0.162$\\ \hline
$10^{-2}$ & $5$ & $85$ & $6.5\times10^{-2}$ & $0.161$\\ \hline
\end{tabular}
\caption{Accuracy and computing time for the S-IEM method}\label{TABLE1}%
\end{table}%

For the case of a nonlocal potential $K$ the division \ of the radial interval
into partitions is not made because the effect of the nonlocal potential would
extend into more than one partition, making the programming more cumbersome.
For the case of a kernel $K(r,r^{\prime}),$ described in Ref. \cite{RPB}, the
accuracy of the S-IEM result \cite{RPB} is also good to $1:10^{-11},$ as is
shown in Fig. 7 of Ref. \cite{RPB}.

For comparison with the S-IEM method some characteristics of the FE-DVR method
are shown in Table \ref{TABLE2}. The potential and the wave number is the same
as in Table \ref{TABLE1}$,$ the radial interval $[0,100\ fm]$ is divided into
$100$ partitions of length of $1\ fm$ each$,$ and the number of Lobatto points
per partition in all partitions is the same but is progressively varied from
$20$ to $7,$ as shown in the table. If one compares the entries in table
\ref{TABLE1} with those in \ref{TABLE2} that correspond to approximately the
same accuracy of $10^{8}$ for tan$(\delta)$ one notices that the FE-DVR method
needs approximately 7 times more support points than the S-IEM$,$ yet the
computing time is between 2 and 3 times less. This remark attests to the
efficiency of the FE-DVR method.%

\begin{table}[tbp] \centering
\begin{tabular}
[c]{|c|c|c|}\hline
$N\ of\ Pts.$ & $err[$tan$(\delta)]$ & $time(s)$\\ \hline \hline
$2000$ & $10^{-10}$ & $0.075$\\ \hline
$1300$ & $10^{-8}$ & $0.050$\\ \hline
$1200$ & $10^{-6}$ & $0.047$\\ \hline
$1000$ & $10^{-4}$ & $0.045$\\ \hline
$700$ & $10^{-2}$ & $0.042$\\ \hline
\end{tabular}
\caption{Accuracy and computing time for the FEM-DVR method}\label{TABLE2}%
\end{table}%

\bigskip

{\Huge Appendix B:\ The round off errors in the FE-DVR method.}

The notation is as follows: $N$ is the number of Lobatto points in each
partition, which is also equal to the number of Lagrange polynomials in each
partition, and $N_{p}$ is the number of partitions. The largest contribution
to the roundoff errors is expected to arise from the solution of Eq. (\ref{5})
for the $N-2$ expansion coefficients. This equation is of the type $\bar
{M}\beta=b,$ where $\beta$ is the column vector of the $N-2$ expansion
coefficients and $\bar{M}$ is a matrix of dimension $N-2,$ whose solution
requires $4\times(N-2)^{3}$ floating point operations. For the case that the
floating point roundoff error of the computer is $\varepsilon$ and that the
errors accumulate linearly, an upper bound for the total error $\varepsilon
_{T}$\ is
\begin{equation}
\varepsilon_{T}\approx4\ast N_{P}(N-2)^{3}\varepsilon. \label{A1}%
\end{equation}
For $N_{p}=100$ and for $\varepsilon=10^{-16},$ which is the value for the
calculations done in MATLAB, one obtains an upper bound for the values of
$\varepsilon_{T}$ that are plotted in Fig. \ref{FIG14} as a function of $N$
\begin{equation}
\varepsilon_{T}\approx4\times10^{-14}(N-2)^{3}\varepsilon. \label{A2}%
\end{equation}

The floating point error that occurs in the calculation of the Lagrange
functions $\ell_{i}(x)$ is much smaller. The numerator contains $N$ factors
$x-x_{i}$ , each of which can be written as $\Delta_{i}+\varepsilon$, where
$\Delta_{i}$ is proportional to the length of each partition. Hence the error
of the product is $\approx \Delta^{N}+N\Delta^{N-1}\varepsilon,$ where $\Delta
$\ is an average value of the $x-x_{i}.$ A similar argument holds for the
denominator and if the error of the numerator adds linearly to the error of
the denominator, then an upper bound for the total error of a Lagrange
function is $\simeq2N\varepsilon/\Delta$. This is much less than the error in
Eq. (\ref{A1}).

For the case of the nonlocal calculation the conditions above are different.
There is only one partition of length $L=15\ fm,$ the number of Lobatto points
is $130$, and the order of each polynomial $\ell(r)$ is $129.$ The error in
the calculation of the Lagrange polynomials, or their derivatives at each
meshpoint, is $\leq \ 2(N-1)\varepsilon/\Delta$. Since the error in the
calculation of the matrix element of $(d^{2}/dr^{2})$ has $N$ terms according
to Eq. (\ref{2d}) that could lead to an error of $\simeq2N(N-1)\varepsilon
/\Delta,$\text{ assuming that all }$\varepsilon$\text{ errors add linearly.
The solution of of Eq. (\ref{5})\ requires }$4(N-2)^{3}$ operations\text{ and
thus an upper bound of the total linear accumulation of }$\varepsilon$ errors
is
\begin{equation}
\simeq2N(N-1)(\varepsilon/L)4(N-2)^{3}=1.8\times10^{-6}. \label{A3}%
\end{equation}
Since the $\varepsilon$ errors do not accumulate linearly, the expected upper
bound for the error could be
\begin{equation}
\leq2(N-1)(\varepsilon/L)4(N-2)^{3}=1.4\times10^{-8}. \label{A4}%
\end{equation}
The above estimate is consistent with the accuracy found numerically in Fig.
\ref{FIG6}.

\medskip{\LARGE Appendix C: Comparison with a finite difference method.}

The finite difference method used for this comparison is Milne's corrector
method, also denoted as the Numerov method, given by Eq. 25.5.21 in Ref.
\cite{AS}. In this method the error of the propagation of the wave function
from two previous points to the next point is of order $h^{6},$ where $h$ is
the radial distance between the consecutive equispaced points. The calculation
is done for the potential $V_{M}$ and for $k=0.5\ fm^{-1}$ as follows.

\begin{figure}
[ptb]
\begin{center}
\includegraphics[
height=2.3601in,
width=3.1384in
]%
{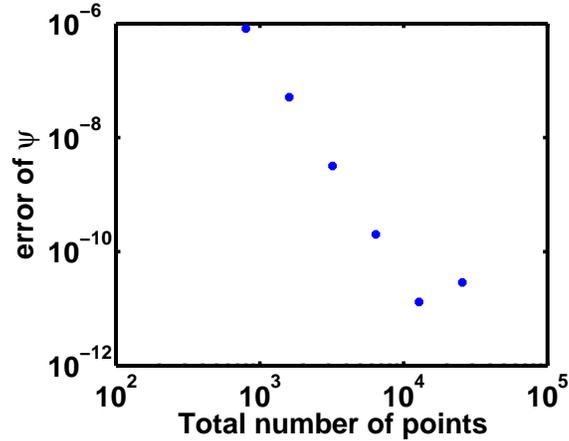}%
\caption{(Color on line) The error of the Numerov wave function at $r=18\ fm,$
as a function of the number $N$ of meshpoints in the interval $[0,20fm]$. The
distance $h$ between points is $20/N$. For each $h$ the wave function is
normalized to the S-IEM wave function at $r=2\ fm.$ The wave number is
$k=0.5\ fm^{-1}$, the potential is $V_{M}.$}%
\label{FIG18}%
\end{center}
\end{figure}

A value of $h$ is selected and the Milne wave function is calculated starting
at the two initial points $r=h$ and $r=2h$ by a power series expansion of the
wave function for the potential $V_{M}$. The values of the wave function for
the additional points $3h,4h,...$are obtained from Milne's method out to the
point $r=20fm.$ The wave function is normalized to the S-IEM value at
$r=2\ fm$, and the error at $r=18\ fm$ is obtained by comparison with the
S-IEM value at that point. The result for a sequence of $h$ values is
illustrated in Fig. \ref{FIG18}.%
For each value of $h$ the wave function is calculated out to $r=100$ $fm$ by
Numerov's method$,$ and the integral (\ref{B10}) is calculated by the extended
Simpson's rule, given by Eq. (25.4.6) in Ref. \cite{AS}. The error is
determined by comparison with the S-IEM result $2.6994702502$ for $\tan
(\delta)$. A comparison with the FE-DVR method is shown in Fig. \ref{FIG24}
and Table \ref{TABLE4} displays the ratio \ Numerov/FE-DVR of the total number
of points and of the time of the two methods for two accuracies of
$\tan(\delta).$ \
\begin{table}[tbp] \centering
\begin{tabular}
[c]{|c|c|c|}\hline
$accuracy\ $tan$(\delta)$ & $ratio\ of\ N\ of\ Pts$ & $time\ ratio$%
\\ \hline \hline
$10^{-6}$ & $\simeq1$ & $20$\\ \hline
$10^{-8}$ & $15$ & $100$\\ \hline
\end{tabular}
\caption{The Numerov/FE-DVR ratio of the required total number of meshpoints and the respective computational times}\label{TABLE4}%
\end{table}%
. More detail of the error and the computing time for the Numerov method is
displayed in Table \ref{TABLE3}.%

\begin{table}[tbp] \centering
\begin{tabular}
[c]{|c|c|c|}\hline
$N\ of\ Pts.$ & $err[$tan$(\delta)]$ & $time(s)$\\ \hline \hline
$12800$ & $1.23\times10^{-9}$ & $51$\\ \hline
$6400$ & $9.41\times10^{-9}$ & $5.8$\\ \hline
$3200$ & $7.50\times10^{-8}$ & $2.1$\\ \hline
$1600$ & $5.99\times10^{-7}$ & $1.0$\\ \hline
$800$ & $4.76\times10^{-6}$ & $0.72$\\ \hline
\end{tabular}
\caption{Accuracy and computing time for the Numerov method}\label{TABLE3}%
\end{table}%
. The calculation is done in MATLAB performed on a desktop using an Intel TM2
Quad, with a CPU Q 9950, a frequency of 2.83 GHz, and a RAM\ of 8 GB.

\bigskip
\end{document}